\documentclass[aps,twocolumn,pra,showpacs,floatfix]{revtex4}
\usepackage{epsfig}
\usepackage{graphicx}
\usepackage{dcolumn}
\usepackage{amssymb,amsmath}
\usepackage{mathrsfs}
\begin{document}

\title{Highly charged ions of heavy actinides as sensitive probes for time variation of the fine structure constant.}

\author{V. A. Dzuba, V. V. Flambaum}

\affiliation{School of Physics, University of New South Wales, Sydney 2052, Australia}

\begin{abstract}
Highly charged ions of heavy actinides from uranium to einsteinium are studied theoretically to find optical transitions sensitive to the variation of the fine structure constant. A number of promising transitions have been found in ions with ionisation degree $\sim$~10. All these transitions correspond in single-electron approximation to the $6p$ - $5f$ transitions. Many of the transitions are between ground and excited metastable states of the ions which means that they can probably be used as optical clock transitions. Some of the ions have more than one clock transition with different sensitivity to the variation of the fine structure constant $\alpha$. The most promising systems include the Np$^{10+}$, Np$^{9+}$, Pu$^{11+}$, Pu$^{10+}$, Pu$^{9+}$, Pu$^{8+}$, Bk$^{15+}$, Cm$^{12+}$, and Es$^{15+}$ ions.

\end{abstract}

\maketitle

\section{Introduction}

Laboratory search for the manifestations of new physics beyond standard model require extremely high accuracy of measurements.
The highest accuracy has been achieved nowadays for the frequencies of optical atomic clocks. Fractional uncertainty of the frequencies measurements gets to the level of $\sim 10^{-18}$ with the prospects of further improvements (see, e.g., reviews \cite{rev1,rev2}). 

Variation of the fine structure constant $\alpha$ may be caused by interaction of scalar or axion dark matter field with electromagnetic field  \cite{Arvanitaki,Stadnik,Stadnik2}. Therefore, measurements of $\alpha$ variation is a method to search for dark matter. Variation of $\alpha$ leads to variation of atomic transition frequencies. 

Therefore, one of the most promising options in search for new physics is monitoring a possible drift of the frequencies of optical clock transitions over long period of time or search for  their oscillating variation. Unfortunately, most of operating optical clocks have low sensitivity  to the variation of $\alpha$~\cite{CJP} (with an exception of Yb$^+$ and Hg$^+$ clocks).
It was suggested in Ref.~\cite{HCI} to use highly charged ions (HCI) in search for sensitive transitions. The sensitivity is proportional to $Z^2(Z_i+1)^2$ where $Z$ is nuclear charge and $Z_i$ is ionisation degree. In single-electron approximation it is also proportional to $\Delta j$, where $j$ is total angular momentum of a single-electron state~\cite{DFW99}
\begin{equation}\label{e:delta}
\Delta_n = \frac{E_n (Z\alpha)^2}{\nu}\left[\frac{1}{j+1/2}-C(j,l)\right],
\end{equation}
where $\Delta_n$ is relativistic energy shift of a single-electron state, $n$ is principal quantum number $E_n$ is ionisation energy from this state, $\nu$ is effective principal quantum number ($E_n=-1/(2\nu^2)$), $C(j,l) \approx 0.6$ is a constant with imitates the many-body effects.

Therefore, the highest sensitivity can be found in $s-f$ or $p-f$ transitions in HCI of heavy atoms (e.g., $\Delta j=3$ for the $s_{1/2} - f_{7/2}$ or $p_{1/2} - f_{7/2}$ transition), i.e., these must be transitions between states of different configurations.
As a rule, the frequencies of such transitions in HCI are well outside of the optical region. To find optical transitions we use the idea of {\em level crossing}~\cite{crossing}.  The energy ordering of the states is different in neutral atoms and hydrogen-like ions. Therefore, if energies are considered as functions of ionisation degree $Z_i$, there must be crossing points at some $Z_i$ for energies of different states. If one of corresponding state is ground state and another is excited metastable state, then the transition between them is likely to be optical transition with features of clock transition with high sensitivity to the variation of the fine structure constant.

A number of promising transitions were suggested in earlier works. Experimental measurements are also in progress (see, e.g., reviews \cite{Safronova,Sahoo}).
The highest sensitivities were found for heavy HCI of actinides Cf$^{16+}$,  Cf$^{17+}$~\cite{Cf-ions}, Cf$^{15+}$, Es$^{16+}$, Es$^{17+}$~\cite{CfEs}, Cm$^{15+}$, Bk$^{16+}$~\cite{CmBk}. There should be  many other  HCI of actinides with optical transitions sensitive to variation of $\alpha$. The aim of present work is to conduct a comprehensive analysis of HCI of heavy actinides from U to Es to find possibly all optical transitions between states of different configurations and identify promising systems which deserve further consideration. We use advanced calculation techniques to study ions with the configuration of external electrons $6s^26p^m5f^n$,  $1 \leq n+m \leq 6$. The number of external electrons varies from three to eight. The transitions between states of different configurations correspond to the $6p-5f$ single-electron transition. 
Majority of found transitions are  $6p_{3/2} - 5f_{5/2}$ or $6p_{3/2} - 5f_{7/2}$ transitions.
They are less sensitive to the variation of $\alpha$ then transitions $6p_{1/2} - 5f_{5/2}$ and  $6p_{1/2} - 5f_{7/2}$ transitions in ions considered before~\cite{Cf-ions,CfEs,CmBk}. However, there are some interesting systems which deserve further consideration.

\section{Method of calculation}

To perform calculations for all ions we use a combination of the linearized single-double coupled cluster (SD) and the configuration interaction methods~\cite{SD+CI}.
The electronic configuration of the ions is $1s^2 \dots 5d^{10}6s^26p^m5f^n$, where $0 \leq n \leq 6$, $0 \leq m \leq 3$ and $1 \leq n+m \leq 6$. 
We consider each ion as a closed-shell core of 78 electrons [$1s^2 \dots 5d^{10}$] plus three to eight external electrons distributed over the $6s$, $6p$ and $5f$ states.
The calculations are performed using the $V^{N-M}$ approximation~\cite{VN-M}, where $N$ is the total number of electrons and $M$ is the number of valence electrons ($M=2+m+n$, $N=78+M$ in our case).
The initial relativistic Hartree-Fock (RHF) procedure is done for the closed-shell core of 78 electrons.
The RHF Hamiltonian includes the Breit interaction and quantum electrodynamic (QED) corrections,
\begin{eqnarray} \label{e:RHF}
	&&\hat H^{\rm RHF}= c\bm{\alpha}\cdot\mathbf{p}+(\beta -1)mc^2+V_{\rm nuc}(r)+ \\
	&&V_{\rm core}(r) + V_{\rm Breit}(r) + V_{\rm QED}(r), \nonumber
\end{eqnarray}
where $c$ is the speed of light, $\bm{\alpha}$ and $\beta$ are the Dirac matrices, $\bm{p}$ is the electron momentum, $m$ is the electron mass, $V_{\rm nuc}$ is the nuclear potential obtained by integrating the Fermi distribution of the nuclear charge density,  $V_{\rm core}(r)$ is the self-consistent RHF potential created by the electrons of the closed-shell core, $V_{\rm Breit}(r)$ is Breit potential~\cite{Breit}, 
$V_{\rm QED}(r)$ is the radiative QED potential~\cite{QED}.
	
After completing the self-consistent procedure for the core, the B-spline technique~\cite{B-spline} is used to create a complete set of single-electron wave functions. The functions are constructed as linear combinations of B-splines, which are eigenstates of the RHF Hamiltonian. We use 40 B-splines of order 9 in a box with a radius of $R_{\rm max}=40a_B$; the orbital angular momentum $0 \leq l \leq 6$.
These basis states are used to solve the SD equations for the core and for the valence states~\cite{SD+CI} and for constructing the many-electron basis states for the CI calculations. 

Solving the SD equations gives us two correlation operators, $\Sigma_1$ and $\Sigma_2$. $\Sigma_1$ describes the correlation interaction between a particular valence electron and the core, whereas $\Sigma_2$ describes the Coulomb interaction screening between a pair of valence electrons~\cite{SD+CI,CI+MBPT}.

The effective CI+SD Hamiltonian has the form
\begin{eqnarray} \label{e:HCI}
	\hat H^{\rm CI}=\sum_{i=1}^{M} \left(\hat H^{\rm RHF}+\Sigma_1\right)_i 
	+\sum_{i<j}^{M} \left(\dfrac{e^2}{|r_i-r_j|}+ \Sigma_{2ij}\right)
\end{eqnarray}
Here $i$ and $j$ enumerate valence electrons, summation goes over valence electrons, $e$ is the electron charge, and $r$ is the position operator of the electrons. 

There is a well-known fact that increasing the number of valence electrons exponentially increases the size of the CI matrix. We have up to eight valence electrons, resulting in a matrix with an exceptionally large size. It will require considerable computational power to handle such matrix. In exchange for a very small loss of accuracy, the size of the CI matrix can be reduced by orders of magnitude by using the CIPT (configuration interaction with perturbation theory) method suggested in~\cite{cipt}. By dividing many-electron basis states into two large groups, low-energy states and high-energy states and ignoring the off-diagonal matrix elements between high-energy states the effective CI Hamiltonian is constructed 
\begin{equation}
	\langle i|H^\mathrm{eff}|j\rangle = \langle i|H^\mathrm{CI}|j\rangle+\sum_{k}\frac{\langle i|H^\mathrm{CI}|k\rangle\langle k|H^\mathrm{CI}|j\rangle}{E-E_{k}}.
	\label{e:HCIPT}
\end{equation}
Here, $H^\mathrm{CI}$ is given by (\ref{e:HCI}), $i, j \leqslant N_{\text {eff }}, N_{\text {eff }}<k \leqslant N_{\text {total}}$, 
$N_{\text {eff }}$ is the number of low-energy states and $N_{\text {total}}$ is the total number of many-electron basis states.
Note that the choice of $N_{\text {eff }}$ is arbitrary. One can increase it until the results become stable. 
Parameter $E$ in (\ref{e:CIPT}) is the energy of the state of interest, and $E_k$ denotes the diagonal matrix element for high-energy states, $E_{k}=\left\langle k\left|H^{\mathrm{CI}}\right| k\right\rangle$. The summation in (\ref{e:HCIPT}) runs over all high-energy states. 

The energies and wave functions are found by solving the matrix eigenvalue problem
\begin{equation}
\left( \mathbf{H}^{\mathrm{eff}} - E\right)\mathbf{X} = 0,
	\label{e:CIPT}
\end{equation}
where $\mathbf{X}$ is the vector of expansion for the many-electron wave function of valence electrons over single-determinant basis states.
Note that the parameter $E$ in the denominator of (\ref{e:HCIPT}) is the same as the energy of the state of interest which is to be obtained from solving the CI equations (\ref{e:CIPT}). Since this energy is not known in advance, iterations over $E$ are needed to find it.
More detailed explanations of the technique can be found in Ref.~\cite{cipt}.

\begin{figure}[tb]
	\epsfig{figure=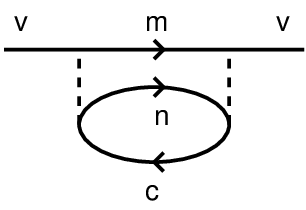,scale=1.0}
	\caption{Diagram representing dominating contribution to the $\Sigma_1$ operator. Index $v$ stands for valence state, $c$ is core state, $m,n$ are virtual states.}
	\label{f:Sigma1}
\end{figure}
For ions with seven or eight external electrons we also use a modification of the method in which the $6s$ electrons are attributed to the core.
This allows to reduce significantly the demand for computer power while having little effect on the CI calculations. When there are many $6p$ and $5f$ electrons, excitation from the $6s$ state are not very important. On the other hand, having the $6s$ state in the core creates a problem on the stage of calculating the one-electron correlation operator $\Sigma_1$. It turns out that the many-body perturbation theory for $\Sigma_1$ does not work due to the presence of terms with small energy denominators. This manifests itself in absence of convergence in solving the SD equations for $s$ or $d$ valence states. Let us consider an example of $\Sigma_1$ operator for $d$ states of Es. The diagram representing dominating term in $\Sigma_1$ is presented on Fig.~\ref{f:Sigma1}.
Corresponding energy denominator is $\epsilon_v+\epsilon_c-\epsilon_n-\epsilon_m$, where $v$ is valence state, $c$ is core state, and $m,n$ are virtual excited states. If we take $\epsilon_v=\epsilon_{5d_{5/2}}$=-11.092~a.u., $\epsilon_c=\epsilon_{6s}$=-15.490~a.u.,$\epsilon_n=\epsilon_m=\epsilon_{6p_{3/2}}$=-13.097~a.u., then $\epsilon_v+\epsilon_c-\epsilon_n-\epsilon_m$=0.388~a.u., which is small and leads to problems. 
One way around this is to take $\epsilon_v=\epsilon_{6p}$ or $\epsilon_v=\epsilon_{5f}$. This is justified because our states of interest have only $6p$ or $5f$ electrons. A more radical way around the problem is to move the $6s$ electrons into valence space. One more reason of doing this is that the problem may manifest itself in calculating the $\Sigma_2$ operator. Note, that the problem of small energy denominators is common for many atoms or ions with $ns^2$ subshell on the border between core and valence states. It may not manifest itself. For example, neutral Tl, with the $6s^26p$ configurations of external electrons can be successfully treated as an atom with one valence electron (see., e.g.~\cite{Ladder}). However, it is important to be aware of the problem.

To calculate the sensitivity of atomic transitions to the variation of $\alpha$ we present the frequencies of the transition in the form
\begin{equation}
\omega(x) = \omega_0+q\left[\left(\frac{\alpha}{\alpha_0}\right)^2-1\right] \equiv \omega_0+qx,
	\label{e:q}
\end{equation}
where index 0 indicate present-day value and $q$ is sensitivity coefficient found from calculations by varying the value of $\alpha$ in compute codes:
\begin{equation}
q=\frac{\omega(+\delta)-\omega(-\delta)}{2\delta}.
	\label{e:dq}
\end{equation}
We use small values of $\delta$ ($10^{-3} \leq q \leq 10^{-2}$) to avoid non-linear effects.
It is also convenient to have the so-called {\em enhancement factors} $K$ ($K=2q/\omega$) which connect the variation of the transition frequencies to the variation of $\alpha$:
\begin{equation}
\frac{\delta(\omega_1/\omega_2)}{(\omega_1/\omega_2)} = \left(K_1-K_2\right)\frac{\delta \alpha}{\alpha}.
	\label{e:K}
\end{equation}
To have high sensitivity to $\alpha$ variation one should measure variation of the ratio of two transition frequencies with large difference in the enhancement factors      $K_1-K_2$.

Accuracy of the calculations depends on the number of external electrons. The most important factor is adequate treatment of inter-electron correlations.
In our approach core-core and core-valence correlations are treated very well within the SD method. Note, that the closed-shell core and corresponding SD equations are the same for all ions of a specific element. This means that the SD equations are to be solved only once for an element and the $\Sigma_1$ and $\Sigma_2$ operators are the same for all ions of the element. In contrast, the valence-valence correlations are treated within the CI approach and the accuracy varies significantly with the number of valence electrons. The uncertainty is on the level of 1\% for the simplest systems with just three valence electrons (including the $6s$ electrons). It increases to about 6\% for systems with five electrons (see more detailed discussion below on the Es$^{16+}$ ion).
Ions with eight external electrons (six if $6s$ electrons are attributed to the core) and open $6p$ and $5f$ subshells have dense spectrum and the uncertainty is larger than the distance between states on the energy scale. However, even for such systems calculations produce a lot of useful information (see more details discussion below on the case of Am$^{9+}$ ion).

\section{Results and discussion}

The search for optical transitions in HCI of heavy actinides is reflected in Table~\ref{t:act}. We study the ions from U$^{11+}$ to Es$^{13+}$.
For each element we consider ions with three to eight external electrons which form configurations from $6s^26p$ or $6s^25f$ to $6s^26p^n5f^m$ ($m+n=6$). 
Thus we are looking for transitions which in single-electron approximation correspond to the $6p$ - $5f$ transition. Table~\ref{t:act} indicates ground states of each ion and first state of a different configuration. The energies of the states are presented in the form $E_J$, where $E$ is the energy in cm$^{-1}$ with respect to the ground state and $J$ is total angular momentum of the state. Dominating configurations are presented in first column. Note that the $6s^2$ subshell is omitted  in the configurations for shorter records. However, the $6s$ electrons are always included into the valence space. Note also that the first state of a different configuration is not necessary the first excited state. There are may be several states below it which belong to the same configuration as the ground state.
The energies of excited states, which are in optical region, are shown in bold. The aim of the table is to identify all ions which have optical transitions between states of different configurations. These ions are further studied in following subsections.

The table indicates that most of the studied transitions are between states of the configurations with more than two $6p$ electrons and less than six $5f$ electrons. This means that the $6p_{1/2}$ subshell is fully occupied and we are dealing with the open $6p_{3/2}$ and $5f_{5/2}$ subshells.
Corresponding transitions in single-electron approximations are the $6p_{3/2} - 5f_{5/2}$ ($\Delta j=1$) transitions, i.e. they are not most sensitive to the variation of $\alpha$ (see formula (\ref{e:delta})). The most sensitive transitions can be found in first two lines of the table. The correspond to the $6p_{1/2} - 5f_{5/2}$ or $6p_{1/2} - 5f_{7/2}$ in which ($\Delta j=2,3$). The sensitivity coefficient $q$ in these transitions is large, $q \sim 4 \times 10^5$~cm$^{-1}$.
Most of these transitions were studied before \cite{Cf-ions,CfEs,CmBk}. The transitions in Bk$^{15+}$ ion have been studied in the present work.

The ions in first two lines of Table~\ref{t:act} have relatively simple electron structure and only a couple of transitions in optical region.
This might be a disadvantage from experimental point of view. In contrast, the ions with more than two $6p$ electrons have more complicated spectra with a number of optical transitions. Some of them have more than one optical transitions which are good candidates for clock transitions and have different sensitivities to the variation of the fine structure constant.

\begin{table*}
\caption{\label{t:act}Transition energies $E_J$ (in cm$^{-1}$) between states of different configurations in HCI of heavy actinides.
All transitions correspond to the $6p - 5f$ single-electron transition. All energies are given with respect to the ground state.
Subscipt $J$ indicates the value of the total angular momentum $J$. Zero energy means ground state. 
Non-zero energy corresponds to the first state of a different configuration. Numbers in bold indicate optical transitions between ground and excited state. 
The dash "$-$" indicates highly excited states ($E_J > 10^5$ cm$^{-1}$). 
$N_v$ is the number of valence electrons above $6s^2$; ionisation degree $Z_i$ is given by $Z_i = Z - N_v - 80$.
All numbers in the table represent results of present work unless indicated otherwise.
Comparison with earlier calculations (where available) are shown in footnotes. }
\begin{ruledtabular}
\begin{tabular}{c l cccccccc}
\multicolumn{1}{c}{$N_v$}&
\multicolumn{1}{c}{Conf.}&
\multicolumn{1}{c}{$_{92}$U}&
\multicolumn{1}{c}{$_{93}$Np}&
\multicolumn{1}{c}{$_{94}$Pu}&
\multicolumn{1}{c}{$_{95}$Am}&
\multicolumn{1}{c}{$_{96}$Cm}&
\multicolumn{1}{c}{$_{97}$Bk}&
\multicolumn{1}{c}{$_{98}$Cf}&
\multicolumn{1}{c}{$_{99}$Es}\\
\hline
1 & 6p &    0$_{1/2}$  & 0$_{1/2}$ & 0$_{1/2}$     & 0$_{1/2}$     & 0$_{1/2}$    &  0$_{1/2}$ & {\bf 23398}$_{1/2}$\footnotemark[2] & 57643$_{1/2}$ \\
  & 5f &    $-$    &    $-$   & 91293$_{5/2}$ & 63190$_{5/2}$ & {\bf 38375$_{5/2}$}\footnotemark[1] & {\bf 8880$_{5/2}$}\footnotemark[1] &  0$_{5/2}$ &  0$_{5/2}$\\ 
&&&&&&&&\\
2 & $6p^2$ &   $0_0$  & $0_0$  & $0_0$ &      $0_0$      &      $0_0$     &  $0_0$                 &  {\bf 7284}$_0$\footnotemark[3]  &  70850$_0$ \\
   & $6p5f$  &      $-$   &     $-$   &    $-$   &   83759$_2$ & 48543$_3$ & {\bf 20715}$_3$   &      0$_3$            & {\bf 7403}$_2$\footnotemark[6] \\
   & $5f^2$   &      $-$  &      $-$   &   $-$    &     $-$            &      $-$         & 65482$_4$          & {\bf 13681}$_4$\footnotemark[4] & 0$_4$ \\
&&&&&&&&\\
3 & $6p^3$     &     $0_{3/2}$     &    $0_{3/2}$              &  {\bf 6853}$_{3/2}$ & 46157$_{3/2}$ & 87135$_{3/2}$ & 94521$_{3/2}$& $-$   & $-$   \\
   & $6p^25f$  & 59418$_{5/2}$ & {\bf 26707}$_{5/2}$  &   $0_{5/2}$          &  $0_{5/2}$        &    $0_{5/2}$      &  $0_{5/2}$   &$0_{5/2}$  & {\bf 24712}$_{9/2}$ \\
   & $6p5f^2$  &       $-$             &        $-$                     &   $-$                    &     $-$              &  86584$_{9/2}$&  81932$_{9/2}$& {\bf 11383}$_{9/2}$\footnotemark[5] & 0$_{9/2}$ \\
    & $5f^3$     &           $-$         &         $-$                   &         $-$                &       $-$             &        $-$             &      $-$     & 48970$_{9/2}$& {\bf 10975}$_{9/2}$\footnotemark[7]  \\
&&&&&&&&\\
4 & $6p^4$     &         $0_2$ &        $0_2$         &      $0_2$           &  $69455_2$        & $-$                    &       $-$            &    $-$     &   $-$  \\
  & $6p^35f$   & 69206$_3$ &  {\bf 39652}$_3$ &  {\bf 5027}$_3$ &  {\bf 16664}$_3$ & 50646$_3$ & $-$ & $-$    & {\bf 13837}$_5$ \\
  & $6p^25f^2$ &      $-$        &             $-$           &          $-$            &       $0_4$           & $0_4$               & $0_4$                &$0_4$ & $0_4$  \\
  & $6p 5f^3$        &        $-$      &            $-$            &            $-$           &           $-$             &        $-$       &  61631$_5$        & {\bf 34722}$_5$ &  $-$  \\
&&&&&&&&\\
5 & $6p^5$ &      $0_{3/2}$       &    $0_{3/2}$     &    $0_{3/2}$              &  {\bf 32200}$_{3/2}$ &         $-$             &         $-$                  &      $-$       &    $-$      \\
   & $6p^45f$ & 66355$_{5/2}$ & 52593$_{5/2}$ & {\bf 8534}$_{5/2}$ &  $0_{5/2}$               & $0_{5/2}$              & {\bf 19671}$_{5/2}$ &     $-$       &    $-$      \\
   & $6p^35f^2$ &           $-$       &            $-$         &    $-$      &  75530$_{7/2}$     & {\bf 37541}$_{7/2}$& 0$_{7/2}$                 &      $-$       & $-$  \\
  & $6p^25f^3$ &           $-$      &             $-$          &             $-$              &            $-$           &       $-$                     &             $-$           &  0$_{9/2}$   & 0$_{9/2}$ \\
  & $6p5f^4$   &       $-$         &              $-$        &             $-$               &                $-$          &       $-$                     &               $-$              &      & 51290$_{9/2}$ \\
&&&&&&&&\\
6 & $6p^6$ &       $0_0$       &        $0_0$   &    $0_0$             &   {\bf 6812}$_0$ & {\bf 3746}$_0$   &            $-$          &   $-$      &   $-$    \\
  & $6p^55f$ &   92144$_1$ &  65544$_1$ & {\bf 15296}$_1$ &     $0_1$            &  {\bf 14656}$_2$ &            $-$        &    $-$    &  $-$      \\
  & $6p^45f^2$ &          $-$     &     $-$           &          $-$             & {\bf 14595}$_0$ &    $0_2$              & {\bf 28336}$_4$ &    $-$    & $-$ \\
  & $6p^35f^3$ &          $-$    &      $-$          &            $-$            &   53650$_0$      &         $-$              &    0$_4$             &  0$_4$   & $-$ \\
  & $6p^25f^4$ &          $-$    &      $-$          &            $-$           &           $-$           &          $-$              &         $-$              &  $-$     & 0$_4$ \\
  
\end{tabular}
\footnotetext[1]{Ref. \cite{CmBk}}			
\footnotetext[2]{ $E$ = 18686 cm$^{-1}$ in Ref. \cite{Cf-ions}.}			
\footnotetext[3]{ $E$ = ~5267 cm$^{-1}$ in Ref. \cite{Cf-ions}.}			
\footnotetext[4]{ $E$ = ~9711 cm$^{-1}$ in Ref. \cite{Cf-ions}.}			
\footnotetext[5]{ $E$ = 12314 cm$^{-1}$ in Ref. \cite{CfEs}.}			
\footnotetext[6]{ $E$ = ~7445 cm$^{-1}$ in Ref. \cite{CfEs}.}			
\footnotetext[7]{ $E$ = 10591 cm$^{-1}$ in Ref. \cite{CfEs}.}			

\end{ruledtabular}
\end{table*}

\subsection{Uranium and neptunium}

\begin{table}
\caption{\label{t:Np} Excitation energies (cm$^{-1}$) of low-lying states of Np ions and their sensitivities to the variation of the fine structure constant.}
\begin{ruledtabular}
\begin{tabular}{l lc rr c}
\multicolumn{1}{c}{Ion}&
\multicolumn{1}{c}{Leading}&&
\multicolumn{3}{c}{This work}\\
&\multicolumn{1}{c}{Configuration}&
\multicolumn{1}{c}{$J$}&
\multicolumn{1}{c}{$E$}&
\multicolumn{1}{c}{$q$}&
\multicolumn{1}{c}{$K$}\\

\hline
Np$^{10+}$ & $6p^3$   & 3/2 &    0  &    0  & 0 \\
          & $6p^25f$ & 5/2 & 26700 & 66000 & 4.9 \\
          & $6p^25f$ & 7/2 & 41400 & 76000 & 3.7 \\

Np$^{9+}$ & $6p^4$   &  2  &    0  &     0 & 0 \\
          & $6p^4$   &  0  & 18535 &  1400 & 0.15 \\
          & $6p^35f$ &  3  & 39652 & 55400 & 2.9 \\
\end{tabular}
\end{ruledtabular}
\end{table}

We start our study from uranium ions. Uranium is attractive in terms of its availability and relatively low radioactivity. Therefore, it is important to check whether its ions have suitable transitions. Unfortunately, we have not found any. We studied the ions from U$^{11+}$ to U$^{6+}$, which have from three to eight external electrons (including $6s$). The minimal $6p-5f$ distance on the energy scale is $\sim 6\times 10^{4}$~cm$^{-1}$ for the $6p^3 \ J=3/2$ to $6p^25f \ J=5/2$
transition in U$^{9+}$. This is outside of the optical region. For other ions the interval is even larger.

The lightest actinides ions which have optical $6p$ - $5f$ transitions are Np$^{10+}$ and Np$^{9+}$ , see Table~\ref{t:Np}. These ions have relatively simple electron structure and only two excited states which are within optical region with respect to the ground state. Both excited states of Np$^{10+}$ ion are not very long living. First is connected to the ground state by a M1 transition. It is suppressed due to domination of different configurations in upper and lower states.
It goes due to configuration mixing and the lifetime of the upper state is $\lesssim$~1~s. The second excited state is connected to the ground state by E2 transition, but it can also decay to a lower state via M1 transition. Its lifetime is $\sim$~0.1~s.

The Np$^{9+}$ ion is more interesting. Its first excited state is connected to the ground state by E2 transition and lifetime is $\sim$~4~s.
Second excited state is connected to the ground state by suppressed M1 transition and its lifetime is likely to be $<$~1~s.
The two transitions have different sensitivity to the variation of the fine structure constant. The comparing of two frequencies can be used for the search of this variation.

\subsection{Plutonium}

\begin{table}
\caption{\label{t:Pu} Excitation energies (cm$^{-1}$) of low-lying states of Pu ions and their sensitivities to the variation of the fine structure constant.}
\begin{ruledtabular}
\begin{tabular}{lr lc rr c}
\multicolumn{1}{c}{Ion}&
\multicolumn{2}{c}{Leading}&&
\multicolumn{3}{c}{This work}\\
&\multicolumn{2}{c}{Configuration}&
\multicolumn{1}{c}{$J$}&
\multicolumn{1}{c}{$E$}&
\multicolumn{1}{c}{$q$}&
\multicolumn{1}{c}{$K$}\\

\hline
Pu$^{11+}$ & 95\% & $6p^25f$ & 5/2 &     0 &      0 & 0 \\      
          & 94\% & $6p^3   $ & 3/2 &  6845 & -72700 & -15 \\
          & 95\% & $6p^25f$ & 7/2 & 16060 &  11900 &  1.5 \\

Pu$^{10+}$ & 82\% & $6p^4   $ &    2 &       0 &       0 &     0 \\
          & 92\% & $6p^3 5f$ &    3 &    5027 &   63700 &    25 \\
          & 82\% & $6p^3 5f$ &    2 &   11545 &   62800 &    11 \\
          & 91\% & $6p^3 5f$ &    4 &   12231 &   67700 &    11 \\
          & 91\% & $6p^3 5f$ &    1 &   15388 &   67600 &    8.8 \\
          & 88\% & $6p^4   $ &    0 &   19878 &   -1800 &  -0.18 \\

Pu$^{9+}$ & 82\% &   $6p^5   $ &   3/2 &      0 &       0 &     0  \\
         & 92\% &   $6p^4 5f$ &   5/2 &  8534 &   80400 &    19 \\
	 & 91\% &   $6p^4 5f$ &   7/2 &  11468 &   78600 &    14 \\
         & 89\% &   $6p^4 5f$ &   3/2 &  11600 &   79700 &    14 \\
         & 90\% &   $6p^4 5f$ &   1/2 &  14214 &   73200 &   10 \\

Pu$^{8+}$ & 88\% &  $6p^6   $ &    0 &       0 &       0 &     0 \\ 
         & 88\% &  $6p^5 5f$ &    1 &   15296 &   89700 &   11 \\
         & 86\% &  $6p^5 5f$ &    2 &   21794 &   87700 &   7.6 \\
         & 76\% &  $6p^5 5f$ &    4 &   23113 &   79600 &    6.6 \\
         & 84\% &  $6p^5 5f$ &    3 &   27178 &   81400 &    5.6 \\
\end{tabular}
\end{ruledtabular}
\end{table}

Plutonium ions have many optical transitions between states of different configurations. The results of calculations for ions from Pu$^{11+}$ to Pu$^{8+}$ are presented in Table~\ref{t:Pu}. Note that we also included the percentage of leading configurations to illustrate the level of configuration mixing.
Many transitions are quite sensitive to the variation of the fine structure constant. However, we do not see good candidates for clock transitions.
Practically all states can decay to lower states via M1 transitions. Many of these transitions are suppressed due to the difference in leading configurations, but not strongly  suppressed because of noticeable  admixture of different configuration, which is clear from the presented percentages.

The values of the sensitivity coefficients $q$ for most of transitions are positive and close in value. This is because the transitions correspond to the transition from lower state $6p$ to upper state $5f$.  The values of enhancement factors $K$ are different mostly due to the difference in the frequency of the transitions ($K=2q/\omega$).

Calculations for Pu$^{11+}$ and Pu$^{10+}$ are done with the $6s$ electrons in the valence space. Calculations for Pu$^{9+}$ and Pu$^{8+}$ are simplified in several ways. First, the $6s$ electrons are moved to the core; only the second-order core-valence correlation operators $\Sigma_1$ and $\Sigma_2$ were used.
Finally, relatively small effective CI matrix, corresponding to few 
 configurations, was used. All other configurations are included perturbatively (see Eq.~(\ref{e:HCIPT})). Consecutively, the accuracy for  the Pu$^{9+}$ and Pu$^{8+}$ ions is lower than for the Pu$^{11+}$ and Pu$^{10+}$ ions. 
The uncertainty is comparable to the distance between energy levels. However, the ground states are established with high level of confidence.

\subsection{Americium}

\begin{table}
\caption{\label{t:Am} Excitation energies (cm$^{-1}$) of low-lying states of Am$^{9+}$ ion and their sensitivities to the variation of the fine structure constant.}
\begin{ruledtabular}
\begin{tabular}{r lc rr c}
\multicolumn{2}{c}{Leading}&&
\multicolumn{3}{c}{This work}\\
\multicolumn{2}{c}{Configuration}&
\multicolumn{1}{c}{$J$}&
\multicolumn{1}{c}{$E$}&
\multicolumn{1}{c}{$q$}&
\multicolumn{1}{c}{$K$}\\

\hline
93\% & $6p^5 5f$   &  1 &       0 &       0 &    0   \\ 
55\% & $6p^5 5f$   &  2 &    2885 &   26900 &    19  \\ 
91\% & $6p^5 5f$   &  4 &    7079 &    1900 &    0.5 \\ 
58\% & $6p^5 5f$   &  3 &    7729 &   23900 &    6.2 \\ 
93\% & $6p^4 5f^2$ &  0 &   10263 &   61100 &    12  \\ 
99\% & $6p^4 5f^2$ &  4 &   11039 &   71900 &    13  \\ 
59\% & $6p^4 5f^2$ &  2 &   11082 &   46500 &    8.4 \\ 
77\% & $6p^4 5f^2$ &  3 &   12608 &   59000 &    9.4 \\ 
95\% & $6p^4 5f^2$ &  1 &   15578 &   71700 &    9.2 \\ 
\end{tabular}
\end{ruledtabular}
\end{table}

The Am$^{11+}$, Am$^{10+}$, and Am$^{9+}$ ions do have optical transitions between states of different configuration (see Table~\ref{t:act}). 
However, corresponding excited states are 
not metastable. They can decay to lower states via partly suppressed M1 transitions.
Since there are many decay channels, the lifetime of these state are not likely to be very long.
The energy interval between states of different configurations goes down with increasing number of valence electrons and thus decreasing of ionisation degree.
This leads to strong configuration mixing for the Am$^{9+}$ ion, which has six electrons in open $6p$ and $5f$ subshells. The spectrum of this ion is dense and calculations for it are difficult. Table~\ref{t:Am} presents the results of sample calculations for the ion. Note however, that the uncertainty of the calculations are larger than the energy interval between states. This means that we cannot guarantee the order of the states nor can we find out real sensitivity to the variation of the fine structure constant. However, there is still a lot of useful information which comes from the calculations:
\begin{itemize}
\item The ground state of the ion is most probably one of the two lowest states of the configurations presented in Table~\ref{t:Am}. The transition between these states is likely to be narrow and sensitive to the variation of the fine structure constant. This comes from the very different mixing of the configurations in the states (see percentage of leading configurations presented in the table), which leads to suppression of the M1 transition and enhancement of the sensitivity to the variation of $\alpha$.
\item It is very likely that there is another metastable state with different sensitivity to the variation of $\alpha$. In present calculations this is state with $E=7079$~cm$^{-1}$ and $J=4$. In reality, it can be some other state.
\end{itemize}
Note that the values of enhancement factor $K$ are not reliable since they strongly depend on the energy of the transition ($K=2q/\omega$).
The values of $q$ are more stable. Note however, that swopping of the states leads to sign change in both, $q$ and $K$. 
In the end we conclude that  the Am$^{9+}$ ion is an interesting system which probably deserves further study with more advanced technique.

\subsection{Curium and Berkelium}

The Cm$^{15+}$ and Bk$^{16+}$ ions were studied before~\cite{CmBk}. The Cm$^{15+}$ ion has only one optical transition, which is the transition between the $6s^26p_{1/2}$ ground and the $6s^25f_{5/2}$ excited states. The transition is actually the $6p_{1/2} - 5f_{5/2}$ transition and it is sensitive to the variation of $\alpha$. The study of other ions did not reveal any promising systems. The  Cm$^{11+}$ and Cm$^{10+}$ ions do have optical transitions between states of different configurations. 
However, they are relatively high in the spectra and have many channels of decay into lower states via M1 or E2 transitions. 
Therefore, corresponding states are not metastable and cannot serve as clock states. 

\begin{table}
\caption{\label{t:Bk} Excitation energies (cm$^{-1}$) of low-lying states of Bk$^{15+}$ ion and their sensitivities to the variation of the fine structure constant.}
\begin{ruledtabular}
\begin{tabular}{r lc rr c}
\multicolumn{2}{c}{Leading}&&
\multicolumn{3}{c}{This work}\\
\multicolumn{2}{c}{Configuration}&
\multicolumn{1}{c}{$J$}&
\multicolumn{1}{c}{$E$}&
\multicolumn{1}{c}{$q$}&
\multicolumn{1}{c}{$K$}\\

\hline
 96\% & $6p^2$  &   0 &     0 &      0 & 0 \\
 98\% & $6p5f$  &   3 & 20715 & 150000 & 14 \\ 
 94\% & $6p5f$  &   2 & 29400 & 196000 & 13 \\
 97\% & $6p5f$  &   3 & 43471 & 205000 & 9.4 \\
\end{tabular}
\end{ruledtabular}
\end{table}
Berkelium ions are more promising. The Bk$^{16+}$ ion has two optical transitions with very high and different sensitivity to the variation of $\alpha$~\cite{CmBk}.
Next ion, Bk$^{15+}$, also has several optical transitions sensitive to the variation of $\alpha$ (see, Table~\ref{t:Bk}). It has at least one very narrow clock transition, which is transition between ground and first excited states ($\Delta J=3$). The leading contribution to the transition probability is mediated by the hyperfine interaction. For example, the magnetic dipole hyperfine interaction mixes states with $J=3$ and $J=2$. This opens the electric quadrupole (E2) transition to the ground state. Next excited state is not metastable since it can decay to the lower state via M1 transition.

\subsection{Californium and Einsteinium}

\begin{table}
\caption{\label{t:Cf} Excitation energies (cm$^{-1}$) and $g$-factors of low-lying states of Cf$^{15+}$ ion.
Comparison with earlier calculations.}
\begin{ruledtabular}
\begin{tabular}{r lc rc rc}
\multicolumn{2}{c}{Leading}&&
\multicolumn{2}{c}{This work}&
\multicolumn{2}{c}{Ref.~\cite{CfEs}}\\
\multicolumn{2}{c}{Configuration}&
\multicolumn{1}{c}{$J$}&
\multicolumn{1}{c}{$E$}&
\multicolumn{1}{c}{$g$}&
\multicolumn{1}{c}{$E$}&
\multicolumn{1}{c}{$g$}\\
\hline
87\% &  $6p^25f$   & 5/2  &      0 &  0.848 &      0 &  0.843 \\
95\% &  $6p5f^2$   & 9/2  &  11386 &  0.813 & 12314 &  0.813 \\
72\% &  $6p^25f$   & 7/2  &  22285 &  1.116 & 21947 &  1.083 \\
89\% &  $6p5f^2$   & 5/2  &  27395 &  0.701 & 26665 &  0.715 \\
89\% &  $6p5f^2$   & 7/2  &  29723 &  0.838 & 27750 &  0.868 \\
91\% &  $6p5f^2$   & 3/2  &  30064 &  0.735 & 28875 &  0.765 \\
\end{tabular}
\end{ruledtabular}
\end{table}

Some of ions of californium and einsteinium were studied before. The Cf$^{17+}$ and Cf$^{16+}$ ions were studied in Ref.~\cite{Cf-ions}; 
the Cf$^{15+}$, Es$^{16+}$ and Es$^{17+}$ ions were studied in Ref.~\cite{CfEs}. Here we present more detailed study of these and other ions of californium and einsteinium. First of all, it is  instructive to compare the results of present and earlier calculations in a non-trivial case of many valence electrons (four of five electrons). 
The calculations were performed in sufficiently different ways so that the comparison of the results reveals important information about the accuracy of both approaches. Both works used the combined SD+CI method~\cite{SD+CI,SD+CIa} but in different implementations. Our earlier work followed Ref.~\cite{SD+CIa} while present work followed Ref.~\cite{SD+CI}.
Different approximations were used in Refs.~\cite{SD+CI} and \cite{SD+CIa} to relate single-electron energies to the many-electron energies of CI states.
Another difference comes from the fact that the $6s$ states were attributed to the core in \cite{CfEs} while they are in valence space in present work.
However, the most important difference is the use of the CIPT technique in the present work to improve the efficiency of the calculations (see Eps.~(\ref{e:HCIPT},\ref{e:CIPT}) above). Neglecting off-diagonal matrix elements between high-energy states allows one to reduce the size of the effective CI matrix and improve the efficiency by many orders of magnitude. 
This comes with the price of some loss in accuracy. It is important to know the value of this loss. Comparing two works is a good way to get some idea about this, since the earlier work~\cite{CfEs} used the full-scale CI calculations.

Table~\ref{t:Cf} compares energy levels and $g$-factors of the Cf$^{15+}$ ion calculated in the present work and in Ref.~\cite{CfEs}.
There is very good agreement between two calculations. The difference in energies is between 1\% and 8\%. 
Good agreement in the $g$-factors indicates very similar composition of the states. If we attribute all the difference in the results to the neglecting of the off-diagonal matrix elements in the CI matrix between high-energy states then few per cent uncertainty in the results is not a big price to pay for many orders of magnitude improvement in the efficiency of the calculations.

Further study of the californium ions with more than two electrons on the $6p$ and $5f$ subshells reveals no more interesting systems.
The Cf$^{14+}$ ion does have an optical transition between states of different configurations. This is a transition between ground state $6s^26p^25f^2 \ J=4$ and excited state $6s^26p5f^3 \ J=5$ at $E=34722$~cm$^{-1}$ (see Table~\ref{t:act}). However, the excited state is relatively high in the spectrum and is not metastable. It has many channels of decay into lower states via the M1 transitions.


\begin{table}
\caption{\label{t:Es} Excitation energies (cm$^{-1}$) of low-lying states of Es ions and their sensitivities to the variation of the fine structure constant}
\begin{ruledtabular}
\begin{tabular}{lr lc rr c r}
\multicolumn{1}{c}{Ion}&
\multicolumn{2}{c}{Leading}&&
\multicolumn{3}{c}{This work}&
\multicolumn{1}{c}{Ref.~\cite{CfEs}}\\
&\multicolumn{2}{c}{Configuration}&
\multicolumn{1}{c}{$J$}&
\multicolumn{1}{c}{$E$}&
\multicolumn{1}{c}{$q$}&
\multicolumn{1}{c}{$K$}&
\multicolumn{1}{c}{$E$}\\

\hline

Es$^{17+}$ &  97\% & $5f^2$ &    4 &     0 &      0 &      &  0 \\
           &  90\% & $5f^2$ &    2 &  7403 & -26000 & -6.5 & 7445\footnotemark[1] \\
           &  98\% & $5f^2$ &    5 & 23652 &  17000 &  1.3 & \\
           &  87\% & $6p5f$ &    3 & 23736 & -159000 & -11 & \\
           &  90\% & $5f^2$ &    4 & 27654 & -14000 & -0.9 & \\
           &  84\% & $5f^2$ &    3 & 29338 & -310000 & -17 & \\

Es$^{16+}$ & 86\% & $6p5f^2$ & 9/2 &     0 &       0 &   0 &      0  \\
           & 56\% & $6p5f^2$ & 5/2 &  6879 & -165000 & -48 &   6994\footnotemark[2]  \\ 
           & 82\% & $5f^3  $ & 9/2 & 10404 &  254000 &  49 &  10591  \\ 
           & 56\% & $6p5f^2$ & 3/2 & 10986 &  120000 &  22 &  11056  \\ 
           & 94\% & $6p5f^2$ & 7/2 & 16368 &   88000 &  11 &  15441  \\ 
           & 43\% & $6p5f^2$ & 5/2 & 23351 & -154000 & -13 &  24301  \\ 
		     	    
Es$^{15+}$ & 96\% & $6p^25f^2$ & 4 &     0 &       0 &   0 &   \\
           & 82\% & $6p^25f^2$ & 2 &  4791 &   51700 &  22 &    \\
           & 96\% & $6p  5f^3$ & 5 &  13837 &  398400 & 57 &    \\
           & 89\% & $6p^2  5f^2$ & 5 & 22554 &  57500 &  5.1 &    \\
           & 54\% & $6p^25f^2$ & 4 & 24278 &  156400 &  13 &    \\
           & 62\% & $6p  5f^3$ & 4 & 33090 &  316400 &  19 &    \\
\end{tabular}
\footnotetext[1]{Ref. \cite{CfEs} gives $q=-46600$ cm$^{-1}$, $K=-13$, leading configuration is $6p5f$.}			
\footnotetext[2]{Ref. \cite{CfEs} gives $q=-184000$ cm$^{-1}$, $K=-53$, leading configuration is the same.}			
\end{ruledtabular}
\end{table}

Table~\ref{t:Es} shows energy levels and sensitivity coefficients ($q$ and $K$, see Eps. (\ref{e:q},\ref{e:K})) calculated in the present and earlier~\cite{CfEs} works.
 In the case of Cf$^{15+}$ the agreement with previous calculations is very good. This illustrates once more the usefulness of the CIPT technique.
It brings huge gain in efficiency while the loss of accuracy is insignificant.

All three Es ions have many transitions with different sensitivity to the variation of the fine structure constant. 
First excited state of each ion is relatively long-living state connected to the ground state by a suppressed M1 transition.

\section{Summary}

We conducted a comprehensive study of HCI of heavy actinides from U to Es with the $6s^26p^m5f^n$ ($1 \leq m+n \leq 6$) configurations of external electrons.
The study was aimed at finding optical transitions which correspond to the $6p-5f$ transitions in single-electron approximation and therefore are sensitive to the variation of the fine structure constant $\alpha$. We also checked which of the transitions have features of atomic clock transitions to ensure high accuracy of the measurement. We identified a number of promising systems in addition to what was considered before. These include the Np$^{10+}$, Np$^{9+}$, Pu$^{11+}$, Pu$^{10+}$, Pu$^{9+}$, Pu$^{8+}$, Bk$^{15+}$, Cm$^{12+}$, and Es$^{15+}$ ions. Some ions studied before (Cf$^{15+}$, Es$^{17+}$, Es$^{16+}$) were studied in more details.
Majority of the considered ions have at least two narrow transitions with different sensitivity to the variation of $\alpha$. The transitions are either E2 or suppressed (due to difference in composition of the states) M1 transitions between the ground and lowest excited states.

\acknowledgments

This work was supported by the Australian Research Council Grants No. DP230101058 and DP200100150.

\end{document}